# Principles of microRNA regulation of a human cellular signaling network

Qinghua Cui[1+], Zhenbao Yu[2+], Enrico O. Purisima[1], and Edwin Wang[1*]

[1]*Computational Chemistry & Biology Group* and [2]*Mammalian Cell Genetics Group, Biotechnology Research Institute, National Research Council Canada, Montreal, Quebec, H4P 2R2, Canada*

[+]These authors contributed equally to this work

[*]To whom correspondence should be addressed. Email: <u>edwin.wang@cnrc-nrc.gc.ca</u>

## Abstract

MicroRNAs (miRNAs) are endogenous ~22-nucleotide RNAs, which suppress gene expression by selectively binding to the 3'-noncoding region of specific message RNAs through base-pairing. Given the diversity and abundance of miRNA targets, miRNAs appear to functionally interact with various components of many cellular networks. By analyzing the interactions between miRNAs and a human cellular signaling network, we found that miRNAs predominantly target positive regulatory motifs, highly connected scaffolds and most downstream network components such as signaling transcription factors, but less frequently target negative regulatory motifs, common components of basic cellular machines and most upstream network components such as ligands. In addition, when an adaptor has potential to recruit more downstream components, these components are more frequently targeted by miRNAs. This work uncovers the principles of miRNA regulation of signal transduction networks and implies a potential function of miRNAs for facilitating robust transitions of cellular response to extracellular signals and maintaining cellular homeostasis.

# Introduction

Recently, a large group of small molecules, called microRNAs (miRNAs) have been found to regulate gene expression by base-paring with target message RNAs (mRNAs), leading to mRNA cleavage or translational repression. As a result, miRNAs control protein concentration at post-transcriptional and translational levels, but do not affect mRNA transcription and protein stability. An accumulating body of evidence reveals that miRNAs have critical functions in various biological processes [Ambros, 2004]. It is currently estimated that miRNAs account for ~ 1% of predicted genes in higher eukaryotic genomes and that up to 10%-30% of genes might be regulated by miRNAs. Furthermore, miRNA targets include signaling proteins, enzymes, and transcription factors (TFs) and so on. The diversity and abundance of miRNA targets offer an enormous level of combinatorial possibilities and suggest that miRNAs and their targets appear to form a complex regulatory network intertwined with other cellular networks such as signal transduction networks. However, it is unclear if and how miRNAs might orchestrate their regulation of cellular signaling networks and how regulation of these networks might contribute to the biological functions of miRNAs. Here we address these questions by analyzing the interactions between miRNAs and a human signaling network.

Eukaryotic cells use complex signaling networks to make decisions about whether to grow, differentiate, move or die [Ferrell, Jr., 2002;Han et al., 2004;Luscombe et al., 2004]. The components of cellular signaling networks, mainly composed of proteins, are activated or inhibited in response to specific input stimuli and in turn serve as stimuli for further downstream proteins. Currently, signaling networks are presented as directed graphs in which nodes represent proteins and links describe qualitative relationships (activation, inhibition or neutral interaction) between proteins. However, the strength of the links or the abundance (concentration) of individual nodes (proteins) in a network also plays an important role in determining signaling strength and specificity. Since the mechanism by which the concentration of signaling proteins is controlled in a living cell is very complicated and poorly understood, a directed and weighted cellular signaling network indicating both direction and strength of links is currently not available. As miRNAs can directly and specifically knock down protein expression, we hypothesize that miRNAs might play an important role in the regulation of the strength and specificity of cellular signaling networks through directly controlling the concentration of network components (proteins) at post-transcriptional and translational levels.

# Results and discussion

## MiRNAs more frequently target network downstream signaling components than ligands and cell surface receptors

To systematically analyze the interactions between miRNAs and signaling networks, we took a literature-mined signaling network which represents signal transduction processes from multiple cell surface receptors to various cellular machines in a mammalian hippocampal CA1 neuron [Ma'ayan et al., 2005]. The network contains 540 nodes and 1,258 links, including 689 activating (positive) links, 306 inhibitory (negative) links and 263 neutral (protein interactions) links (Supplementary Figure 1). Most of the signaling pathways in the network represent commonly used pathways in many cell types [Ma'ayan et al., 2005]. Hence, we expect the conclusions in this study could be transferable to other cell types.

To uncover which signaling proteins are potentially regulated by miRNAs in the network, we took genome-wide computationally predicted miRNA target genes from two recent studies [Krek et al., 2005; Lewis et al., 2005]. These targets were predicted based on the principle of miRNA-target interactions. Over 70% (4,431) of the miRNA targets predicted in the two studies are overlapped. The accuracy of the prediction methods has been confirmed by experimental validation of randomly selected targets [Stark et al., 2005] and by large-scale gene expression profiling studies [Lim et al., 2005]. Up to 88% of randomly selected predicted miRNA targets were proved to be real targets [Stark et al., 2005]. We mapped all the overlapped miRNA targets onto the 540 network proteins. We found that 159 (29.4%) of the network proteins are miRNA targets, while the miRNA targets represent only ~17% of total genes in human genome ($P < 2 \times 10^{-4}$). This result indicates that miRNAs more frequently target signaling proteins than others and implies that miRNAs may play a relatively more important role in regulating signaling networks than in other cellular processes.

To discover which stages of the signal information flow are predominantly regulated by miRNAs, we first sorted the network components into several groups, e.g. ligands, cell surface receptors, intracellular signaling proteins and nuclear proteins, based on their positions in the signaling information flow, and mapped miRNA targets onto each group. We found that the fraction of miRNA targets in a group (the ratio of the number of the miRNA targets to total number of proteins in the group) increases with the signal information flow from the upstream to the downstream (Figure 1). For example, only 9.1% of the ligands are miRNA targets, whereas half of the nuclear proteins, most of which are TFs, are miRNA targets. In other words, the miRNA targets are enriched more than five times in the most downstream group compared to the most upstream group. A similar result was obtained when we extended this analysis to the signaling proteins in a genome scale (see Supplementary Information).

## MiRNAs preferentially target the downstream components of the adaptors, which have potential to recruit more downstream components

Many intracellular signaling activities such as recruiting downstream signaling components to the vicinity of receptors are performed by adaptor proteins. Adaptors do not have enzyme activity, but physically interact with upstream and downstream signaling proteins. They activate, inhibit or relocalize downstream components through direct protein-protein interactions. One adaptor is able to recruit distinct downstream components in different cellular conditions [Wu et al., 2006]. Accordingly, if an adaptor can recruit more downstream components, these components should have a higher dynamic gene expression behavior. Because miRNAs are known to have a high spatio-temporal expression behavior, we asked if miRNAs preferably regulate the downstream components of the adaptors, which have potential to recruit more downstream components. To answer this question, we first extracted downstream components binding to the network adaptors. We then counted the number of the downstream components for each adaptor and divided the adaptors into two groups. One group contains adaptors linking to four or less downstream components (low-link group), while the other group contains adaptors linking to more than four downstream components (high-link group). Each group contains a similar number of downstream adaptor-binding components (91 and 108, respectively). We found that the fraction of miRNA targets in the high-link group (36.1%, 39/108) is obviously higher than that in the low-link group (24.2%, 22/91, 0.015). This result suggests that miRNAs preferentially target the downstream components of the adaptors, which have potential to recruit more downstream components. For example, the adaptor Grb2 directly interacts with 14 downstream signaling proteins, half of which are miRNA targets (Figure 2). These downstream components are functionally involved in different signaling pathways that lead to different cellular outputs. For example, SHC regulates cell growth and apoptosis through activation of small GTPases of the Ras family, while NWASP is involved in the regulation of actin-based cytoskeleton through activation of small GTPases of the Rho family. These two components are targeted by different miRNAs. To accurately respond to extracellular stimuli, adaptors need to selectively recruit downstream components. As miRNA targets are enriched in the downstream components of high-linked adaptors, miRNAs may play an important role for precise selection of cellular responses to stimuli by controlling the concentration of adaptors' downstream components. In contrast, no correlation between the abundance of miRNA targets in the upstream components and the link number of the adaptors to the upstream components (Supplementary Table I). In addition, we found no correlation between the abundance of miRNA targets of adaptors and their link numbers.

**MiRNAs more frequently target positively linked network motifs**
A complex signaling network can be broken down into distinct regulatory patterns, or network motifs, typically comprised of three to four interacting components capable of signal processing [Babu et al., 2004;Barabasi and Oltvai, 2004]. Network motifs, which represent the simplest building blocks, are statistically overrepresented subgraphs in a network [Kharchenko et al., 2005;Lee et al., 2002;Shen-Orr et al., 2002;Teichmann and Babu, 2004;Wuchty et al., 2003]. Each of them bears a distinct regulatory function in cellular networks. The function of a motif also depends on whether the links are positive or negative. For example, positive feedback loops lean to emergent network properties such as ultrasensitivity, bistability and switch-like behavior, while negative feedback loops perform adaptation, desensitization, and preservation of homeostasis [Balazsi et al., 2005;Dekel et al., 2005;Ferrell, Jr., 2002;Luscombe et al., 2004]. In our previous work, we showed that mRNA decay plays an important role in motif regulatory behavior [Wang and Purisima, 2005]. Thus, analyzing how miRNAs interact with network motifs may shed some insights into understanding miRNA regulation principles in signaling networks. We identified 11 types of motifs in the network (Table I, Supplementary Figure 2). We mapped miRNA targets onto the network motif nodes, and classified each type of motif into several subgroups based on the number of nodes that are miRNA targets. For example, the three node network may have none of their nodes as a miRNA target (category 0), or may have just one of their nodes as a miRNA target (category 1), or 2 (category 2) or all three as miRNA targets (category 3). For each motif, we calculated the ratio of positive links to the total directional (positive and negative) links (termed as Ra) in each subgroup and compared it with the average Ra in all the motifs, which is shown as a horizontal line in Figure 3. For most motifs except Motifs id46, id204 and id904, the Ra in the subgroup in which none of the nodes are miRNA targets is less than the average Ra of all the motifs (Figure 3, $P < 4 \times 10^{-3}$, Wilcoxon Ranksum test). This result suggests that miRNAs less frequently target negative regulatory motifs. In contrast, for most motifs except Motifs id46 and id4546 which show very low abundance in the network (Table I), the preponderance of positive links in the subgroups increased as the number of miRNA-targeted components rose (Figure 3 and Table I). These motifs, especially the three most abundant motif types, bifan motif (Motif id204), four-node feed forward motif (Motif id904), and Motif id460, show a clear positive correlation between positive link ratio and miRNA target number in the motif. More significantly, when all nodes are miRNA targets in a motif, all the links in the motif are positive links ($P < 0.01$, Wilcoxon Ranksum test, Table I). These results suggest that miRNAs have high potential to target positively linked motifs. For example, AP1 (activator protein 1), CREB (cAMP-responsive element-binding protein) and CBP (CREB-binding protein) form a three-node positive feedback loop. All of the three proteins are miRNA targets.

We then turned to a special three-node regulatory motif built-up by one scaffold protein neutrally linked to other two proteins that are either positively or negatively connected (Motif id110). Scaffold proteins exert their functions through protein-protein interactions. Unlike adaptors, scaffold proteins do not directly activate or inhibit other proteins but provide regional organization for activation or inhibition between other proteins. We found 32 such motifs built-up by 23 scaffold proteins, 6 of which are miRNA targets. These 6 miRNA-targeted scaffold proteins, e.g. CRK, PSD95, SAM68, SHC, SNAP25 and YOTIAO, form 11 scaffold motifs. On average, each miRNA-targeted scaffold protein forms almost two motifs (11/6), a level much higher than that (21/17 or 1.24) formed by the scaffold proteins which are not miRNA targets ($P < 0.05$). In contrast, we found no correlation between the abundance of miRNA targets and the link numbers of the whole network nodes. These results suggest that highly linked scaffold proteins have higher probability to be targeted by miRNAs.

Scaffold proteins are able to recruit distinct sets of proteins to different pathways and thus maintain the specificity of signal information flows. Higher linked scaffold proteins can recruit more protein sets and have a higher degree of spatio-temporal expression behavior. Since the expression of miRNAs is highly specific for tissues and developmental stages, it might be expected that higher linked scaffold proteins would be regulated by more miRNAs. To test this hypothesis, we checked the miRNAs that target the highly linked scaffold proteins. Indeed, we found that each of these scaffold proteins is targeted by several different miRNAs ($P < 0.01$, Supplementary Table II). For example, CRK and SNAP25 are targeted by six miRNAs (miR-1, miR-10a, miR-126,

miR-133a, miR-20 and miR-93) and five miRNAs (miR-1, miR-128a, miR-130a, miR-153 and miR-27b), respectively.

Motifs do not exist in isolation but are embedded into larger subgraphs. Network themes are examples of such larger subgraphs which are enriched topological patterns containing clusters of overlapping motifs, represent a higher order of regulatory relationships between signaling proteins and tie to particular biological functions [Zhang et al., 2005]. To get insights into the higher order regulatory relationships between signaling proteins that are regulated by each miRNA, we explored the network themes. Interestingly, for most of miRNAs, the network motifs containing the targets of a single miRNA form 1 or 2 network themes with a size of more than 20 nodes. Furthermore, most of the themes are associated with one or more of the five cellular machines: transcription machinery, translation machinery, secretion apparatus, motility machinery and electrical response (see Supplementary Information).

**MiRNAs avoid targeting common components of cellular machines in the network**
The signaling network could lead to activation of five distinct cellular machines that include transcription machinery, translation machinery, secretion apparatus, motility machinery and electrical response. Networks contain many functional modules that may carry out specific functions. In cellular signaling networks, functional modules represent a set of proteins that are always present in various cellular conditions. The shortest path from an input node to an output node allowed us to identify such functional modules. To obtain the proteins for a functional module, we extracted the shortest path proteins from each receptor (input node) to all output nodes of each cellular machine. We then examined the fractions of miRNA targets in each cellular machine. We found that the fractions of miRNA targets in almost all functional modules are significantly lower than that (159/540=0.294) in the network (Supplementary Table III). This result indicates that the shortest path proteins have less chance to be miRNA targets.

We then asked if miRNAs preferentially interact with the common components shared by the shortest paths leading to these cellular machines. To answer this question, we extracted proteins found in the shortest path from each cell surface receptor to each output signal protein in each cellular machine. We identified 70 proteins that are shared by all of the five cellular machines in the shortest paths. We found that only 14.3% of them are miRNA targets (Table II), a significant under-representation compared to the fraction of miRNA targets (29.4%) in the network ($P < 2 \times 10^{-4}$). This result suggests that miRNAs avoid disturbing basic cellular processes, because these common proteins are highly shared by basic cellular machines and should be frequently used in various cellular conditions.

**Sensitivity analysis**
Although the accuracy of the miRNA target prediction methods has been well demonstrated by experimental validation of randomly selected targets, 12% of them could not be proved as real targets. To test the potential effects of the errors, we performed the sensitivity analysis as described previously [Wuchty et al., 2003]. We mimicked false positives by randomly adding extra 10% and 20% of network proteins, which are not predicted miRNA targets, to the target list, performed the same analysis and recalculated the P values. In addition, we also removed 10% and 20% of miRNA targets to determine the effect of false negatives. As shown in Supporting Tables IV, V, VI and VII, the results and P values indicate that the trend remains unchanged by the addition of the false positives or false negatives (see more details in Supplementary Text). Therefore, the results we obtained are robust against substantial errors.

**Conclusion remarks**
In this study, we analyzed the distribution of miRNA targets in the cellular signaling network at different levels, ranging from local structures (network motifs) to the network in a global scale (information flow, proteins with distinct biochemical features, and common proteins of basic cellular machines). We found that miRNAs preferentially regulate positive regulatory motifs, highly connected scaffolds and downstream network components such as TFs. In addition, when an adaptor has potential to recruit more downstream components, these components are more frequently regulated by miRNAs. These results revealed that miRNAs regulate signaling networks in multiple ways. By selectively regulating positive regulatory motifs, highly connected scaffolds and the most network downstream components, miRNAs may provide a mechanism to terminate the preexisting messages and facilitate quick and robust transitions for responses to new signals. These functions fit the spatio-temporal behavior of miRNA expression. On the other hand, miRNAs less frequently target negative regulatory motifs, common proteins of basic cellular machines and upstream network components such as ligands. Further discussion of the results is presented in Supplementary Information. Our analysis provides systems-level insights into the interactions between miRNAs and signaling regulatory networks and generates a set of testable hypotheses, which are helpful for understanding the principles of the regulation of signaling networks by miRNAs.

## Materials and methods

### Datasets used in this study

We took a literature-mined signaling network, which represents signal transductions from multiple cell surface receptors to various intracellular machines in a mammalian hippocampal CA1 neuron [Ma'ayan et al., 2005]. After carefully examining the dataset, we noted that five duplicated components were mistakenly included in the network. We confirmed this finding by communicating with Avi Ma'ayan, the first author of the original paper. We merged these duplicated components and finally got a network containing 540 nodes and 1,258 links, in which there are 689 activating links, 306 inhibitory links and 263 neutral links (Supplementary Text files 1 and 2).

We took genome-wide computationally predicted miRNA target genes from two recent studies [Krek et al., 2005;Lewis et al., 2005] . We found that over 70% (4,431) of the miRNA targets predicted by Lewis et al. overlap with these predicted by Krek et al. The overlapped miRNA targets were saved in Supplementary Text file 3 and used in this study.

### Mapping miRNA targets onto the network

We mapped miRNA targets onto the network proteins by writing a Java program (Supplementary Java code file 1). The miRNA-targeting proteins and non miRNA targeting proteins are saved in Supplementary Text files 4 and 5, respectively. To get a global view of miRNA target distribution, we created a Pajek format file (Supplementary Text file 6) and generated a network (Supplementary Figure 1).

### Sorting the network nodes (proteins) along the signal information flow

We sorted the network nodes into four subgroups: ligands, cell surface receptors, cytosolic and organelle (such as mitochondria, ribosomes and vesicles) proteins, and nuclear proteins based on their locations in the signaling flow, and then mapped miRNA targets onto each subgroup. The components of each subgroup were listed in Supplementary Text file 7. We then calculated the fractions of miRNA targets in each subgroup.

### Extracting adaptors and the up- and down-stream components of the adaptors

We extracted adaptors and their directly-linked up- and down- stream components from the network (Supplementary Text file 8). Based on the link numbers of the adaptors to the downstream components, we divided the adaptors into two groups. One group contains adaptors linking to four or less downstream components (low-link group), the other contains adaptors linking to more than four downstream components (high-link group). The downstream components of both groups were listed in Supplementary Text file 9. We mapped miRNA targets to each group and calculated the fractions of miRNA targets.

### Analyzing miRNA target abundance in network motifs

We used Mfinder program [Kashtan et al., 2004] to extract network motifs. The input data for finding network motifs are in Supplementary Text file 10. We list all 3- and 4-size network motifs in Supplementary Text files 11, 12, 13 and 14, respectively. For further analysis of these motifs, we extracted all the members of each type of motif from the network and saved them in compressed Zip files (Supplementary Zip files 1 and 2).

We then counted the number of miRNA targets in each motif and classified each type of motif into several subgroups based on the number of nodes that are miRNA targets. We used the Java source code file (Supplementary Java file 2) to perform sub-grouping. The results are saved in Supplementary Zip files 3 and 4 for 3- and 4- size motifs, respectively. We then calculated the ratio (Ra) of the activation links (+) to the total activation and inhibitory links (+ and -) in each subgroup.

**Sorting different cellular machines of the network**

At the most downstream of the network, the network flow reaches to the output nodes of five cellular machines, such as transcription machinery, motility machinery and so on. The network output nodes in each cellular machine were sorted by Avi Ma'ayan et al. and presented in the Supplementary Information of the original publication [Ma'ayan et al., 2005]. We extracted the subsets of the output proteins (nodes) of each cellular machine from the network (Supplementary Zip file 5). We then extracted the shortest path proteins from each input node (receptor) to all output nodes of each cellular machine using Supplementary Java file 3, which implemented Dijkstra's algorithm. Proteins in each functional module (a collection of distinct shortest path proteins from receptors to the machinery output nodes in a cellular machinery) were saved in Supplementary Zip file 6. After removing input and output nodes of each cellular machine from the shortest path proteins (Supplementary Zip file 7), we extracted the common proteins that are present in every function module (Supplementary Text file 15). We then mapped miRNA targets to each functional module and the common proteins (Supplementary Text file 16), and calculated the fractions of miRNA targets.

**Randomization tests**

To test the statistical significance of observations, we performed randomization tests. A more detailed explanation of randomization tests was previously described by Wang and Purisima [Wang and Purisima, 2005].

## Supporting data and Java source code files

See Sup_1.pdf and SupportingDataFiles.zip

## Acknowledgements

We thank Drs. D. Morse, M. Whiteway and C. Wu for comments on the manuscript. This research is partially supported by Genome Health Initiative.

Reference List

Ambros V (2004) The functions of animal microRNAs. *Nature* **431**: 350-355

Babu MM, Luscombe NM, Aravind L, Gerstein M, Teichmann SA (2004) Structure and evolution of transcriptional regulatory networks. *Curr Opin Struct Biol* **14**: 283-291

Balazsi G, Barabasi AL, Oltvai ZN (2005) Topological units of environmental signal processing in the transcriptional regulatory network of Escherichia coli. *Proc Natl Acad Sci U S A* **102**: 7841-7846

Barabasi AL, Oltvai ZN (2004) Network biology: understanding the cell's functional organization. *Nat Rev Genet* **5**: 101-113

Dekel E, Mangan S, Alon U (2005) Environmental selection of the feed-forward loop circuit in gene-regulation networks. *Phys Biol* **2**: 81-88

Ferrell JE, Jr. (2002) Self-perpetuating states in signal transduction: positive feedback, double-negative feedback and bistability. *Curr Opin Cell Biol* **14**: 140-148

Han JD, Bertin N, Hao T, Goldberg DS, Berriz GF, Zhang LV, Dupuy D, Walhout AJ, Cusick ME, Roth FP, Vidal M (2004) Evidence for dynamically organized modularity in the yeast protein-protein interaction network. *Nature* **430**: 88-93

Kashtan N, Itzkovitz S, Milo R, Alon U (2004) Efficient sampling algorithm for estimating subgraph concentrations and detecting network motifs. *Bioinformatics* **20**: 1746-1758

Kharchenko P, Church GM, Vitkup D (2005) Expression dynamics of a cellular metabolic network. *Molecular System Biology* **doi:10.1038/msb4100023**:

Krek A, Grun D, Poy MN, Wolf R, Rosenberg L, Epstein EJ, MacMenamin P, da P, I, Gunsalus KC, Stoffel M, Rajewsky N (2005) Combinatorial microRNA target predictions. *Nat Genet* **37**: 495-500

Lee TI, Rinaldi NJ, Robert F, Odom DT, Bar-Joseph Z, Gerber GK, Hannett NM, Harbison CT, Thompson CM, Simon I, Zeitlinger J, Jennings EG, Murray HL, Gordon DB, Ren B, Wyrick JJ, Tagne JB, Volkert TL, Fraenkel E, Gifford DK, Young RA (2002) Transcriptional regulatory networks in Saccharomyces cerevisiae. *Science* **298**: 799-804

Lewis BP, Burge CB, Bartel DP (2005) Conserved seed pairing, often flanked by adenosines, indicates that thousands of human genes are microRNA targets. *Cell* **120**: 15-20

Lim LP, Lau NC, Garrett-Engele P, Grimson A, Schelter JM, Castle J, Bartel DP, Linsley PS, Johnson JM (2005) Microarray analysis shows that some microRNAs downregulate large numbers of target mRNAs. *Nature* **433**: 769-773

Luscombe NM, Babu MM, Yu H, Snyder M, Teichmann SA, Gerstein M (2004) Genomic analysis of regulatory network dynamics reveals large topological changes. *Nature* **431**: 308-312

Ma'ayan A, Jenkins SL, Neves S, Hasseldine A, Grace E, Dubin-Thaler B, Eungdamrong NJ, Weng G, Ram PT, Rice JJ, Kershenbaum A, Stolovitzky GA, Blitzer RD, Iyengar R (2005) Formation of regulatory patterns during signal propagation in a Mammalian cellular network. *Science* **309**: 1078-1083

Shen-Orr SS, Milo R, Mangan S, Alon U (2002) Network motifs in the transcriptional regulation network of Escherichia coli. *Nat Genet* **31**: 64-68

Stark A, Brennecke J, Bushati N, Russell RB, Cohen SM (2005) Animal MicroRNAs confer robustness to gene expression and have a significant impact on 3'UTR evolution. *Cell* **123**: 1133-1146

Teichmann SA, Babu MM (2004) Gene regulatory network growth by duplication. *Nat Genet* **36**: 492-496

Wang E, Purisima E (2005) Network motifs are enriched with transcription factors whose transcripts have short half-lives. *Trends Genet* **21**: 492-495

Wu C, Jansen G, Zhang J, Thomas DY, Whiteway M (2006) Adaptor protein Ste50p links the Ste11p MEKK to the HOG pathway through plasma membrane association. *Genes Dev* **20**: 734-746

Wuchty S, Oltvai ZN, Barabasi AL (2003) Evolutionary conservation of motif constituents in the yeast protein interaction network. *Nat Genet* **35**: 176-179

Zhang LV, King OD, Wong SL, Goldberg DS, Tong AH, Lesage G, Andrews B, Bussey H, Boone C, Roth FP (2005) Motifs, themes and thematic maps of an integrated Saccharomyces cerevisiae interaction network. *J Biol* **4**: 6

**Table I Relations between the abundance of positive links with the abundance of miRNA targets in each network motif**

| Motif ID | Ratio of positive links to total positive and negative links in each subgroup | | | | |
|---|---|---|---|---|---|
| | 0 target | 1 target | 2 targets | 3 targets | 4 targets |
| **46** | 10/10 | 11/20 | 2/2 | * | * |
| **98** | 28/48 | 13/24 | 1/3 | 3/3 | * |
| **108** | 12/30 | 11/28 | 9/14 | * | * |
| **110** | 9/15 | 10/11 | 5/5 | 1/1 | * |
| **204** | 1624/2180 | 987/1436 | 226/372 | 85/108 | 12/12 |
| **460** | 148/258 | 79/135 | 29/42 | 10/12 | 18/18 |
| **904** | 403/552 | 265/380 | 163/220 | 55/68 | 4/4 |
| **972** | 6/12 | 40/76 | 21/30 | 2/2 | * |
| **4546** | 42/65 | 29/40 | 5/10 | * | * |
| **4556** | 39/62 | 56/204 | 18/32 | 11/12 | * |
| **5068** | 2/7 | 9/18 | 9/19 | 4/4 | 1/1 |

**Table II Abundance of miRNA targets in the common nodes that locate in the shortest paths of all five cellular machines**

| Cellular machines | Transcription | Translation | Secretion apparatus | Motility | Electrical response |
|---|---|---|---|---|---|
| Number of nodes | 142 | 94 | 115 | 161 | 130 |
| Number of overlapped nodes | 70 | | | | |
| Number of miRNA targets | 10 | | | | |
| P value | $< 2 \times 10^{-4}$ | | | | |

Note: the input nodes (receptors) and output nodes of cellular machines were not counted. The P value was given by comparing the abundance of miRNA targets of the common nodes to that of miRNA targets (0.294) in the network using a randomization test.

**Figure legends:**

**Figure 1 Distribution of miRNA targets in the signal network at different signaling stages.** Signaling proteins are divided into four groups, e.g. ligands, cell surface receptors, intracellular central signaling proteins and nuclear proteins based on their cellular locations in the signaling pathways. The P values were obtained by comparing the fractions of miRNA targets between each group and the whole network using randomization tests.

**Figure 2 MiRNAs preferentially target the downstream components of high-link adaptors.** (A) and (B) illustrate high- and low- link adaptors and their downstream components, respectively. MiRNA targets are in red.

**Figure 3 Relations between the fractions of positive links and the fractions of miRNA targets in network motifs.**
Each type of motif was classified into several subgroups based on the number of nodes that are miRNA targets. For example, the three node network may have none of their nodes as a miRNA target (category 0), or may have just one of their nodes as a miRNA target (category 1), or 2 (category 2) or all three as miRNA targets (category 3). The ratio of positive links to total positive and negative links in each subgroup was plotted as function of miRNA target numbers per motif. The horizontal lines indicate the ratio of positive links to the total positive and negative links in all of the respective network motifs. The network motif ID numbering system is from Alon's motif dictionary (http://www.weizmann.ac.il/mcb/UriAlon/NetworkMotifsSW/mfinder/motifDictionary.pdf).

Figure 1

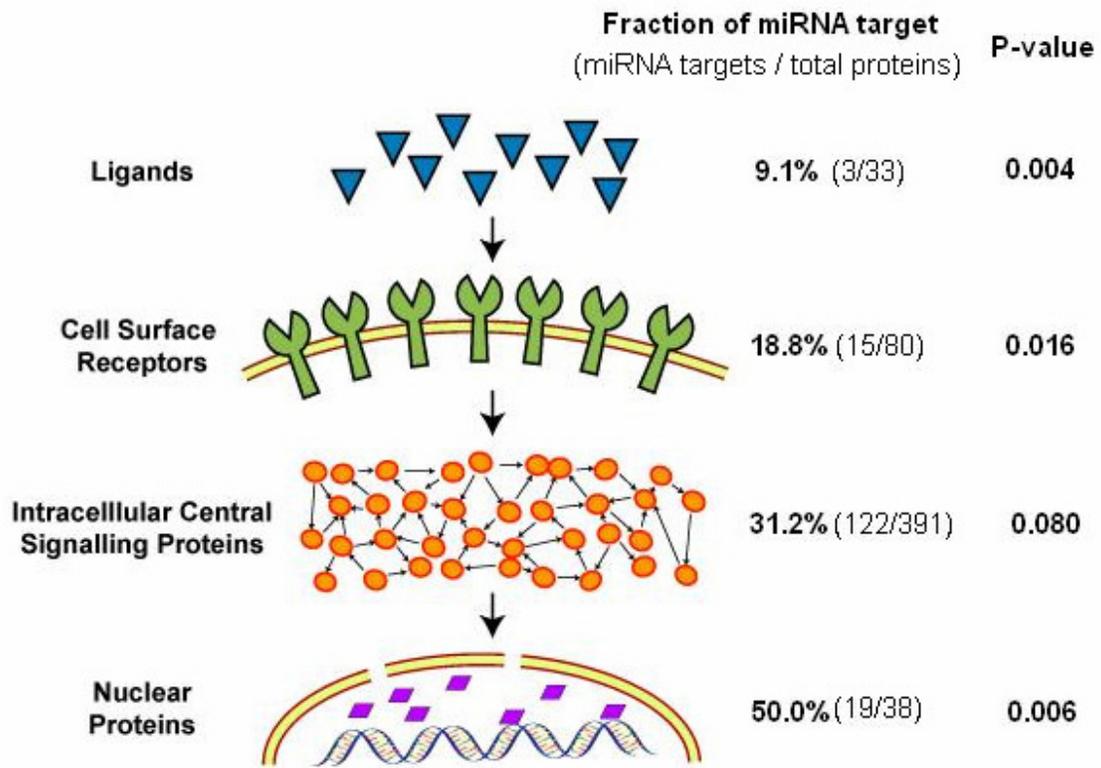

Figure 2

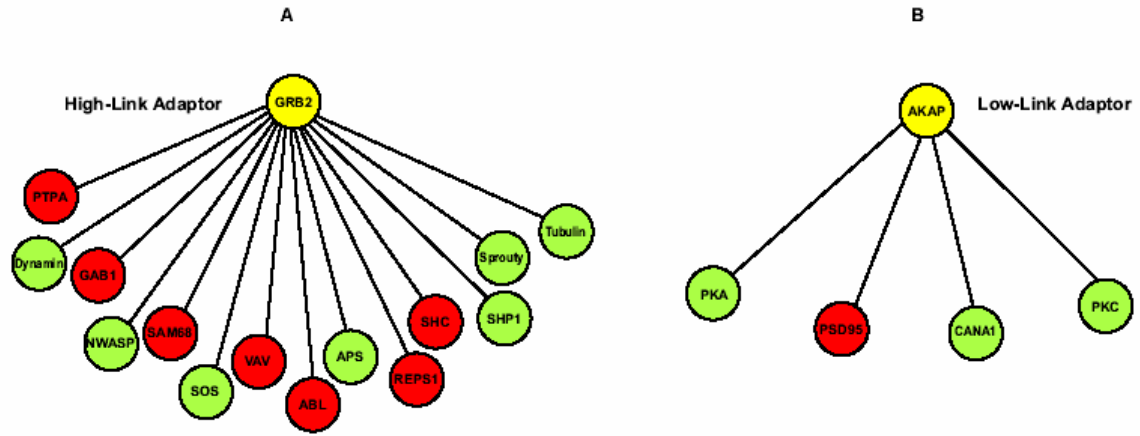

Figure 3

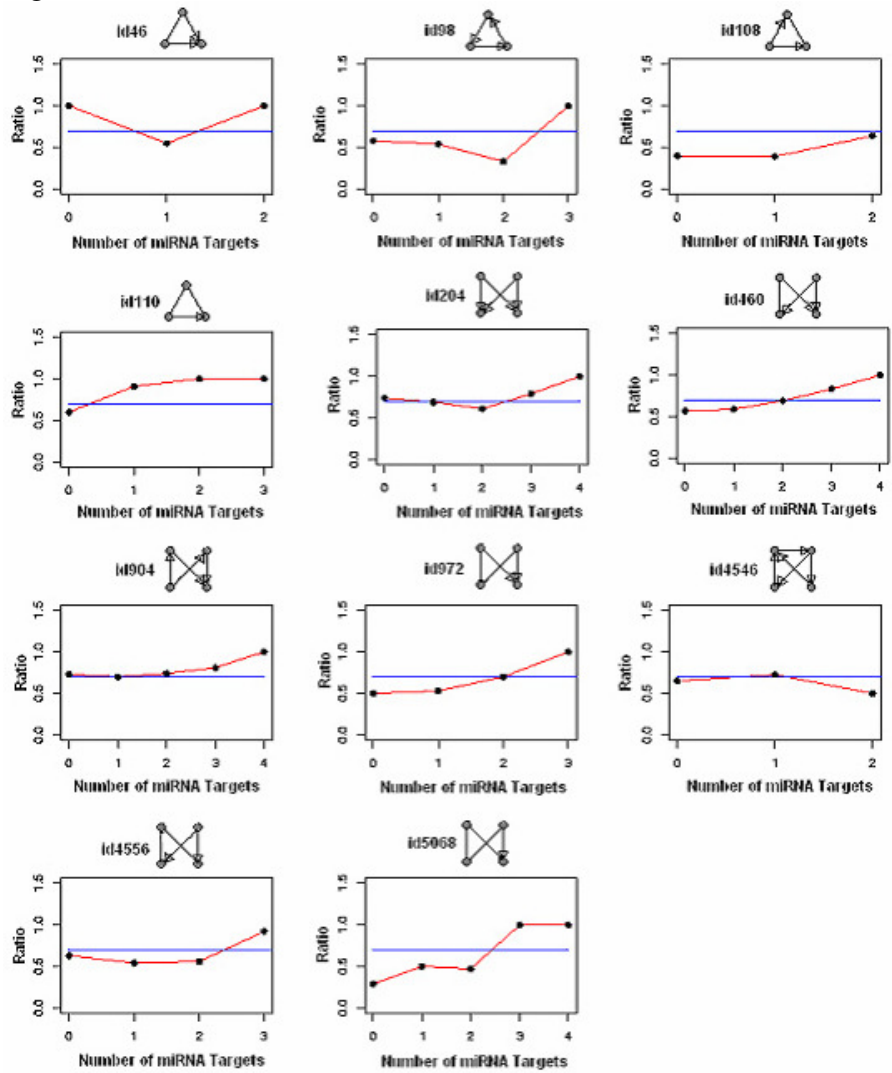